\documentclass[12pt]{article}
\usepackage{amsmath}
\usepackage{amssymb}
\begin{document}
\vskip 1truein
\begin{center}
{\bf SELFDUAL SPIN 2 THEORY IN A 2+1 DIMENSIONAL
\\CONSTANT CURVATURE SPACE-TIME  }
 \vskip 5pt
{\bf P\'{\i}o J. Arias}${}^{a, }${\footnote {e-mail:
parias@fisica.ciens.ucv.ve}} {\bf and} {\bf Rolando Gaitan
D.}${}^{b,
}${\footnote {e-mail: rgaitan@uc.edu.ve}} \\
${}^a${\it Centro de F\'{\i}sica Te\'orica y Computacional,
Facultad de Ciencias, \\Universidad Central de Venezuela, Caracas
1041-A, Venezuela.}\\
${}^b${\it Departamento de F\'\i sica, Facultad de Ciencias y
Tecnolog\'\i a, Universidad de Carabobo, A.P. 129
Valencia 2001, Edo. Carabobo, Venezuela.}\\
\end{center}
\vskip .5truein
\begin{abstract}
The Lagrangian constraint analysis of the selfdual massive spin 2
theory in a 2+1 dimensional flat space-time and its extension to a
curved one, are performed. Demanding consistence of degrees of
freedom in the model with gravitational interaction, gives rise to
physical restrictions on non minimal coupling terms and
background. Finally, a constant curvature scenario is explored,
showing the existence of forbidden mass values. Causality in these
spaces is discussed. Aspects related with the construction of the
reduced action and the one-particle exchange amplitude, are noted.

\end{abstract}

\vskip .5truein

In the context of ordinary field theory, there has been great
interest about the lagrangian study of higher spin fields with
external interaction\cite{ArDe}-\cite{BGP}. These theories are
only known in certain backgrounds (i.e., constant curvature, non
Einstein's spaces), because, in general, a consistent higher spin
field theory with interaction does not exist as a result of the no
conservation of the degrees of freedom and causality violation.

It is well known\cite{FiPa,CSH} that the introduction of auxiliary
fields, that vanish on shell is needed in order to obtain a
lagrangian formulation, in a flat space-time, for a massive field
with spin $s$, without interaction, described by a symmetric,
transverse and traceless, rank $s$, tensor (i.e., ${\partial}^\mu
h_{\mu {\mu _1}...{\mu _{s-1}}} =0, \,\,\, {h^\mu }_{\mu {\mu
_1}...{\mu _{s-2}}} =0$). However, when an arbitrary interaction
is turned on, auxiliary fields become dynamic and hence they could
modify the number of local degrees of freedom.

Causality violation has also been noted\cite{BGP,VeZ,Zwa}. For
this, let us use the following notation\cite{BGP}: The equations
of motion for an integer spin field, $h_{\alpha _1 \alpha _2...}$,
which come from some lagrangian formulation, can be written as
$({\mathcal{M}_{\beta _1 \beta _2...\alpha _1 \alpha _2
...}})^{\mu \nu } {\nabla}_{\mu }{\nabla}_{\nu }h^{\alpha _1
\alpha _2...}
 +... =0$, with the help of the lagrangian constraints\cite{GTy}.
Let the vector $n_\mu$, used to define the characteristic matrix
$\mathcal{M}_{AB}(n)\equiv {\mathcal{M}_{AB}}^{\mu\nu}n_\mu
n_\mu$, where $A,B$ are composed indexes. The solutions of the
characteristic equation $\det{\mathcal{M}_{AB}(n)}=0$, define
characteristic surfaces that might describe some propagation
process. If the solution of the characteristic equation gives rise
to a real $n_0$, the system is called hyperbolic. An hyperbolic
system is called causal if there is no time like vectors among the
solutions of the characteristic equation (on the contrary, if
there exists time like vectors, the corresponding characteristic
surfaces are space like and violate causality). When an arbitrary
external interaction is considered, the characteristic matrix,
$\mathcal{M}_{AB}(n)$ does not necessarily define an
hyperbolic-causal equation of motion system.

In this work we are interested in the study of some dinstinctive
features related to the aforementioned problems in the lagrangian
formulation of the selfdual massive spin 2 field in $2+1$
dimensions\cite{AD,DeserAD}, coupled with gravity\cite{G}. This
letter is organized as follows. We will start with a brief review
of the lagrangian constraints analysis of selfdual massive spin 2
theory in a flat space-time without external interaction. Next, we
introduce the coupling between selfdual massive spin 2 field with
an arbitrary gravitational background, through a suitable set of
non minimal terms in the lagrangian formulation, and we will
discuss the physical restrictions that arise in order to preserve
a consistent interaction. As it is expected, one can find a
constant curvature space solution, in which the degrees of freedom
must be consistently preserved and causality must take place.
There, we obtain forbidden mass values for the selfdual massive
spin 2 field. The construction of the reduced action in constant
curvature space-time is noted. At the end we discuss the
one-particle exchange amplitude. Finally, some remarks will be
stated.

%============================================================================================

The action of selfdual massive spin 2 field\cite{AD}, in flat
space-time is
\begin{equation}
S_{sd}= \int \frac{d^3 x}{2} (m\,\epsilon ^{\mu \nu \lambda
}{h_\mu }^\alpha
\partial _\nu
h_{\lambda \alpha }-m^2\,(h_{\mu \nu }h^{\nu \mu }-h^2 )) \, \, ,
\label{eq1}
\end{equation}
where $h={h^\mu}_\mu$, $\epsilon ^{012} \equiv \epsilon ^{12}
=+1$, and Minkowski's metric is $diag(-++)$. Equation of motion
coming from $S_{sd}$, provide nine primary constraints
\begin{equation}
\phi ^{(1)\mu \rho} =m\,\epsilon ^{\mu \nu \lambda }\partial _\nu
{h_\lambda}^\rho + m^2(\eta ^{\mu \rho }h - h^{\rho \mu }) \approx
0 \, \, . \label{eq2}
\end{equation}
Preservation of (\ref{eq2}), take us to the secondary
constraints
\begin{equation}
\phi ^{(2)\rho} \equiv {\dot \phi}^{(1) 0 \rho } \approx \partial
_\mu \phi ^{(1)\mu \rho } \equiv m^2 \partial ^\rho h -
m^2\partial _\mu h^{\rho \mu }
 \approx 0
\, \, . \label{eq3}
\end{equation}
We observe that (\ref{eq3}) can be replaced with the combination $
\phi ^{(2)\rho } \approx
\partial _\mu \phi ^{(1)\mu \rho} - m\epsilon ^{\rho \mu \alpha }
{\phi ^{(1)}}_{\mu \alpha  } \equiv -m^3 \epsilon ^{\rho \mu
\alpha } h_{\mu \alpha } \approx 0$, which enforces $h_{\mu \nu}$
to be symmetric. Relations, ${\dot \phi}^{(1) i \rho } =0$ allow
us to find the following accelerations
\begin{equation}
{\ddot h}_{k \rho } = \partial _k {\dot h}_{0 \rho } + m\epsilon
_{i k }({\delta ^i }_ \rho  {\dot h}- {{\dot h}_\rho }\,^i ) \, \,
, \label{eq4}
\end{equation}
and the ${\ddot h}_{0 \rho }$ remain unknown.

Procedure continues with the preservation of $\phi ^{(2)\rho} \approx
0$, that gives rise to three additional constraints
\begin{equation}
\phi ^{(3)\rho} \equiv {\dot \phi}^{(2) \rho } \approx -m^3
\epsilon ^{\rho \mu \alpha } {\dot h}_{\mu \alpha }
 \approx 0
\, \, , \label{eq5}
\end{equation}
saying that the symmetry property is consistent with time
evolution. If we look at (\ref{eq5}), the $\rho =0$ component can
be rewritten, on shell, as $ \phi ^{(3) 0} \approx \partial _\rho
\phi ^{(2)\rho } \equiv -m^3 \epsilon ^{\rho \mu \alpha }
\partial _\rho h_{\mu \alpha } \approx 2 m^4 h \approx 0 $,
which shows the traceless property of the tensor field. Then,
preserving $\phi ^{(3)\rho} \approx 0$ we obtain the last
constraint
\begin{equation}
\phi ^{(4)} \equiv 2 m^4 \dot h \approx 0 \, \, , \label{eq6}
\end{equation}
and two relations for the remaining accelerations, $-m^3 \epsilon
^{k \mu \alpha } {\ddot h}_{\mu \alpha }
 = 0$. This allow us to obtain
\begin{equation}
{\ddot h}_{0 k }= {\ddot h}_{ k 0}=\partial _k {\dot h}_{00 } +
m\epsilon _{i k }({\delta ^i }_0  {\dot h}- {{\dot h}_0 }\,^i )\,
\, . \label{eq7}
\end{equation}
The analysis of the lagrangian constraints ends whith the
preservation of (\ref{eq6}). This provides one more relation for
the accelerations, $m^4 \ddot h = 0$ from which we obtain
\begin{equation}
{\ddot h}_{00}= -{\ddot h}_{ ii}= -\partial _k {\dot h}_{0 k } -
m\epsilon _{i k }({\delta ^i }_k  {\dot h}- {{\dot h}_k }\,^i )\,
\, . \label{eq8}
\end{equation}
So, it can be shown that the $16$ lagrangian constraints indicate
the existence of one propagated excitation, and it is described by
a symmetric, transverse and traceless tensor field. In other words
\begin{equation}
{h^{(s)Tt}}_{\mu \nu}={h^{(s)Tt}}_{\nu \mu}\,\,,\,\, \,\,\partial
^{\mu}{h^{(s)Tt}}_{\mu \nu}=0\,
\,,\,\,\,\,{{h^{(s)Tt}}_{\mu}}^{\mu}=0\,\, \, , \label{eq9}
\end{equation}
respectively, satisfying the field equation
\begin{equation}
\epsilon ^{\mu \nu \lambda }\partial _\nu
 {{h^{(s)Tt}}_\lambda }^\rho - mh^{(s)Tt\rho \mu}=0  \, \, .
\label{eq10}
\end{equation}
A Klein-Gordon type equation, $(\Box - m^2){h^{(s)Tt}}_{\mu \nu}=0
$ is obtained from (\ref{eq10}) using (\ref{eq9}).

It can be observed that restrictions (\ref{eq9}) leave just two
free components of the nine in $h_{\mu\nu}$, but relying to
dynamic restriction (\ref{eq10}), it can be seen that only one
degree of freedom is locally propagated. From the action point of
view, one can also expose this unique excitation through the
construction of the reduced action ($S^*_{sd}$), which starts
performing a $2+1$ splitting for $h_{\mu\nu}$, this means,
$n=h_{00} ,\, \,N_i=h_{i0} , \,\,M_i=h_{0i},\,\,
{h^{(s)}}_{ij}=\frac{1}{2}(h_{ij}+h_{ji}) ,
\,\,V=\frac{1}{2}\,\epsilon _{ij}h_{ij}$, in action
(\ref{eq1})\cite{PJA}. Then, a transverse-longitudinal
decomposition is realized introducing new variables defined by:
$N_i \equiv \epsilon _{ik}\partial _k N^T +
\partial _i N^L $ , \,$M_i \equiv \epsilon _{ik}\partial _k M^T +
\partial _i M^L $ , \,$ {h^{(s)}}_{ij} \equiv (\delta _{ij}\Delta
-\partial _i \partial _j)h^{TT} +\partial _i
\partial _j h^{LL} +( \epsilon _{ik}\partial _k
\partial _j+\epsilon _{jk}\partial _k \partial _i )h^{TL}$. This
decomposition establishes an easy way to obtain the reduced action
using the corresponding field equations,
$S^*_{sd}=\int d^3 x \{P\dot{Q}-\frac12 P^2 + \frac12 \,Q(\Delta
-m^2)Q\}$, where $Q\equiv \sqrt 2 \, \Delta h^{TT}$ and $ P\equiv
\sqrt 2 \,m \Delta h^{TL}$, which describes a single massive mode.

%============================================================================================

Now we outline the model of selfdual massive spin 2 field non
minimally coupled with gravity in a torsionless space-time\cite{G}
as follows
\begin{equation}
S_{sdg}=\int \frac{d^3 x}{2} \sqrt {-g}( m\,\varepsilon ^{\mu \nu
\lambda }{h_\mu }^\alpha \nabla _\nu h_{\lambda \alpha }+
{\Omega}^{\alpha \beta \sigma \lambda }h_{\alpha \beta  }h_{\sigma
\lambda } ) \, \, , \label{eq13}
\end{equation}
where $\nabla _\nu $ is the covariant derivative and $ \varepsilon
^{\mu \nu \lambda } \equiv \frac {\epsilon ^{\mu \nu \lambda }}
{\sqrt {-g}}$. Due to the fact that in $2+1$ dimensions the
Riemann curvature tensor can be written in terms of the Ricci
tensor (i.e., $R_{\lambda \mu\nu \rho}=g_{\lambda\nu}R_{\mu
\rho}-g_{\lambda\rho}R_{\mu\nu}-g_{\mu\nu}R_{\lambda\rho}+g_{\mu\rho}R_{\lambda\nu}-\frac
R2 \,(g_{\lambda\nu}g_{\mu \rho}-g_{\lambda\rho}g_{\mu\nu}) $), so
the non minimal coupling in (\ref{eq13}) is characterized by a
tensor ${\Omega}^{\alpha \beta \sigma \lambda }$, whose general
form is
\begin{eqnarray}
{\Omega}^{\alpha \beta \sigma \lambda } \equiv m^2(g^{\sigma
\lambda }g^{\alpha \beta }-
    g^{\sigma \beta }g^{\alpha \lambda})+ a_1 (R^{\sigma \lambda }g^{\alpha \beta
}+
    R^{\alpha \beta }g^{\sigma \lambda})+ a_2 (R^{\sigma \beta }g^{\alpha
\lambda } + R^{\alpha \lambda }g^{\sigma \beta})\nonumber \\ + a_3
R^{\alpha \sigma }g^{\beta \lambda } + a_4 R^{\beta \lambda
}g^{\alpha \sigma } +  a_5 Rg^{\alpha \beta }g^{\sigma \lambda} +
a_6 Rg^{\sigma \beta }g^{\alpha \lambda } +  a_7 Rg^{\lambda \beta
}g^{\sigma \alpha }
 \, \, , \label{eq14}
\end{eqnarray}
with the property ${\Omega}^{\alpha \beta \sigma \lambda }
={\Omega}^{\sigma \lambda\alpha \beta  } $ and real parameters
$a_n$, $n=1,...,7 $.

Taking arbitrary variations on $h_{\mu\nu}$ in $S_{sdg}$ gives
rise to the following field equations
\begin{equation}
\Phi ^{(1) \mu \alpha }\equiv m\,\varepsilon ^{\mu \nu \lambda }
\nabla _\nu {h_\lambda}^\alpha + \Omega ^{\mu  \alpha \sigma
\lambda  } h_{\sigma \lambda } \approx 0 \, \, , \label{eq15}
\end{equation}
wich constitute nine primary constraints.

Three more constraints arise when  $ \Phi ^{(1) o \rho }\approx 0$
is preserved
\begin{eqnarray}
\Phi ^{(2) \alpha }\approx \nabla _\mu  \Phi ^{(1) \mu \alpha
}\equiv \Omega ^{\mu \alpha \sigma \lambda }\nabla _\mu h_{\sigma
\lambda}+\mathcal{B} ^{\alpha \sigma \lambda }h_{\sigma \lambda}
\approx 0\, \, , \label{eq16}
\end{eqnarray}
where
\begin{eqnarray}
\mathcal{B} ^{\alpha\sigma \lambda }\equiv \frac m2 \,\varepsilon
^{\mu \nu \rho}({R^{\alpha \lambda}}_{\nu \mu}\,{\delta
^{\sigma}}_{\rho} -{R^{\sigma}}_{\rho \nu \mu}\,g^{\alpha
\lambda}) +\nabla _\mu \Omega ^{\mu \alpha \sigma \lambda }\, \, .
\label{eq17}
\end{eqnarray}

On the other hand, preservation of $ \Phi ^{(1) k \alpha }\approx
0$ leads to six relations for the accelerations ($m\neq 0$, as in
the flat case)
\begin{eqnarray}
{\nabla _0}^2 {h_j}^\alpha=-\frac{\varepsilon_{0kj}}{m}\,\Omega
^{k \alpha \sigma \lambda }\nabla _0 h_{\sigma
\lambda}-\big(\frac{\varepsilon_{0kj}}{m}\,\nabla _0\Omega ^{k
\alpha \sigma \lambda }+{R^\sigma}_{ 0j0}\,g^{\alpha
\lambda}\big)h_{\sigma \lambda}+\nonumber \\
+ \,\nabla _j\nabla _0{h_0}^\alpha +{R^{\alpha\lambda}}_{
j0}\,h_{0\lambda}\,\,
 , \label{eq18}
\end{eqnarray}
remaining the unknown accelerations,  ${\nabla _0}^2 h_{0 \lambda
}$.

At this point, the lagrangian analysis with free coupling
parameters in arbitrary background is equivalent to the flat
space-time case. The next step is the preservation of the
constraint $\Phi ^{(2) \alpha } \approx 0$, which leads to
\begin{eqnarray}
\Phi ^{(3) \alpha } \equiv \nabla _0 \Phi ^{(2) \alpha }\approx
{\Omega}^{0\alpha 0\lambda}{\nabla _0}^2 h_{0 \lambda}+
\big(\Omega ^{0 \alpha j \lambda }+\Omega ^{j \alpha 0 \lambda }
\big)\nabla _j\nabla _0 h_{0 \lambda}+ \nonumber \\
+\,\big(-\frac{\varepsilon_{0kl}}{m}\,\Omega ^{0 \alpha l \rho
}{\Omega ^{\mu \lambda k}}_\rho +\nabla_0 \Omega ^{0 \alpha \mu
\lambda}+\mathcal{B} ^{\alpha \mu \lambda } \big)\nabla _0
h_{\mu\lambda}+ \nonumber \\
+\big[ -\frac{\varepsilon_{0kl}}{m}\,\Omega ^{0 \alpha l \rho
}\nabla_0{\Omega ^{\mu \lambda k}}_\rho - \Omega ^{0 \alpha l
\lambda}{R^\mu}_{0 l0} + \nabla_0\mathcal{B} ^{\alpha \mu \lambda
}+ \nonumber \\-\Omega ^{\nu \alpha \mu \rho }{R^\lambda}_{\rho
\nu 0} -\Omega ^{\sigma \alpha \nu \lambda }{R^\mu}_{\nu \sigma
0}\big]h_{\mu\lambda}-\Omega ^{0 \alpha l \rho }{R^\lambda}_{\rho
l0}h_{0\lambda}+\nonumber
\\ +\nabla_0 \Omega ^{i
\alpha \mu \lambda}\nabla _i h_{\mu \lambda} + \Omega ^{i \alpha j
\lambda}\nabla _i\nabla_0 h_{j \lambda}\approx 0
  \,
\, , \label{eq19}
\end{eqnarray}
and we expect that this expression represents three additional
constraints, as in the flat case. But, from (\ref{eq19}) it would
be impossible to obtain any relation for the accelerations
${\nabla _0}^2 h_{0 \lambda }$, because (\ref{eq19}) constitutes a
complete system for the aforementioned accelerations. We demand
that all matrixes $3\times 3$, $2\times 2$ and $1\times 1$ built
with ${\Omega}^{0\alpha 0\lambda}$, have null determinant (i.e.,
${\Omega}^{0\alpha 0\lambda}$ totally degenerated), in other words
\begin{equation}
\Omega ^{0 \alpha 0 \lambda } =0 \, \, . \label{eq20}
\end{equation}
This condition gives rise to restrictions on coupling parameters.
Using (\ref{eq14})
\begin{eqnarray}
a_1 = -a_2 \equiv a \,\,, \,\, a_6 = -a_5 \equiv b \,\, , \,\, a_3
= a_4 = a_7 = 0 \,\,, \label{eq21}
\end{eqnarray}
and just two free parameters remain. Then,
\begin{eqnarray}
{\Omega}^{\alpha \beta \sigma \lambda } = a\, R^{\alpha \sigma
\beta \lambda } +
 (m^2 +(\frac{a}{2}-b )R) (g^{\alpha \beta }g^{\sigma \lambda}
- g^{\sigma \beta }g^{\alpha \lambda } )
 \, \, , \label{eq22}
\end{eqnarray}
and now
\begin{eqnarray}
{\Omega}^{\alpha \beta \sigma \lambda } ={\Omega}^{\sigma \lambda
\alpha \beta }=-{\Omega}^{\alpha \lambda\sigma \beta  }
 \, \, . \label{eq23}
\end{eqnarray}
The object $\mathcal{B} ^{\alpha\sigma \lambda }$, can be
rewritten in terms of the Einstein's tensor,
$G_{\mu\nu}=R_{\mu\nu}-\frac{g_{\mu\nu}}{2}\,R$ as follows
\begin{eqnarray}
\mathcal{B} ^{\alpha\sigma \lambda }\equiv  m\,\varepsilon
^{\alpha \lambda \beta}\,{G^{\sigma}}_{\beta} +\nabla _\mu {\Omega
^{\mu \alpha \sigma \lambda }}=-\mathcal{B} ^{\lambda\sigma \alpha
}\, \, ,\label{eq24}
\end{eqnarray}
with an antisymmetric property in virtue of (\ref{eq23}).

With the help of (\ref{eq22}) we can write the three
constraints,
 $\Phi ^{(3) \rho } \equiv \nabla _0 \Phi ^{(2) \rho }
\approx 0$ as follows
\begin{eqnarray}
\Phi ^{(3) \alpha } = \mathcal{N}^{\alpha \lambda }\nabla _0h_{0
\lambda} +\nabla _0\Omega ^{ i \alpha 0 \lambda} \nabla _ih_{0
\lambda}+\mathcal{A}^{ \alpha \lambda} h_{0 \lambda}+\Omega ^{ i
\alpha j \lambda} \nabla _i \nabla _0h_{j \lambda}\nonumber
\\+\,\mathcal{C}^{
\alpha j \lambda} \nabla _0h_{j \lambda}+\nabla _0\Omega ^{ i
\alpha j \lambda} \nabla _i h_{j \lambda}+\mathcal{D}^{ \alpha j
\lambda} h_{j \lambda}
 \approx 0 \,
\, , \label{eq25}
\end{eqnarray}
where
\begin{eqnarray}
\mathcal{N}^{\alpha \lambda } \equiv \frac 1m
\,\varepsilon_{0kl}\,\Omega ^{0 \alpha k \rho}{\Omega ^{0 \lambda
l}}_\rho + \mathcal{B}^{\alpha 0 \lambda}=-\mathcal{N}^{\lambda
\alpha }  \, \, , \label{eq26}
\end{eqnarray}
\begin{eqnarray}
\mathcal{A}^{\alpha \lambda } \equiv -\frac 1m \,\Omega ^{0 \alpha
l \rho}\big(\varepsilon_{0kl}\,\nabla_0{\Omega ^{0 \lambda
k}}_\rho +m {R^\lambda}_{ \rho l 0}+m {R^0}_{ 0 l
0}{\delta^\lambda}_\rho \big)\nonumber \\ +\,
\nabla_0\mathcal{B}^{\alpha 0 \lambda}-\Omega ^{\mu \alpha 0
\rho}{R^\lambda}_{ \rho \mu 0} -\Omega ^{\mu \alpha \nu
\lambda}{R^0}_{ \nu \mu 0}\, \, , \label{eq26a}
\end{eqnarray}
\begin{eqnarray}
\mathcal{C}^{\alpha j \lambda } \equiv -\frac 1m
\,\varepsilon_{0kl}\,\Omega ^{0 \alpha l \rho}{\Omega ^{j \lambda
k}}_\rho +\mathcal{B}^{\alpha j \lambda}+ \nabla_0\Omega ^{0
\alpha j \lambda}\, \, , \label{eq26b}
\end{eqnarray}
\begin{eqnarray}
\mathcal{D}^{\alpha j \lambda } \equiv -\frac 1m
\,\varepsilon_{0kl}\,\Omega ^{0 \alpha l \rho}\nabla_0{\Omega ^{j
\lambda k}}_\rho +\nabla_0\mathcal{B}^{\alpha j \lambda}-\Omega
^{\mu \alpha j \rho}{R^\lambda}_{ \rho \mu 0}\nonumber \\-\Omega
^{\mu \alpha \nu \lambda}{R^j}_{ \nu \mu 0}-\Omega ^{0 \alpha l
\lambda}{R^j}_{ 0l 0}\, \, . \label{eq26c}
\end{eqnarray}

Going on with the lagrangian procedure, preservation of $\Phi
^{(3) \rho } \approx 0$ must represents, as in the flat case, two
expressions for accelerations ${\nabla _0}^2 h_{0 \sigma }$ and
one for the last constraint (whose preservation allow us to get
the remaining accelerations, and the procedure ends). Let us
consider $3\times 3$ and $2\times 2$ arrays  built with objects
$\mathcal{N}^{\alpha \lambda }$, the last request means that
\begin{eqnarray}
\det (\mathcal{N}^{\alpha \lambda } )=0 \,\,, \label{eq27}
\end{eqnarray}
 \begin{eqnarray}
\det (\mathcal{N}^{ij } ) \neq 0 \,\,. \label{eq28}
\end{eqnarray}
Relation (\ref{eq27}) due to the antisymmetry property of the odd
rank matrix $(\mathcal{N}^{\alpha \lambda } )$is identically
satisfied. (\ref{eq28}) gives a physical restriction on the
gravitational field, and it conduces to
\begin{eqnarray}
\varepsilon_{0ij}\,\mathcal{N}^{ij }  \neq 0 \,\,. \label{eq29}
\end{eqnarray}
It can be shown that this restriction, that should be satisfied in
order to keep consistence in the number of degrees of freedom,
will include non Einsteinian solutions. Although these type of
solutions exist, (\ref{eq29}) will impose conditions on them. For
illustration, let consider $R_{\lambda \mu\nu \rho}=\frac{f(x)}{6}
\,(g_{\lambda\nu}g_{\mu \rho}-g_{\lambda\rho}g_{\mu\nu})$. Then,
restriction (\ref{eq29}) enforce a constraint for $f(x)$ (i.e.,
$6M^4 - m^2f(x)+m\sigma {\varepsilon ^k}_0
\partial _k f(x)\neq 0$, with $\sigma
\equiv \frac{2}{3}\,a - b$). Our interest is focussed in the
particular solution $\partial_\mu f(x)=0$ (hence (\ref{eq29})
relates the mass with cosmological constant), which is of dS/AdS
type.

Considering a constant curvature space-time, with cosmological
constant $\lambda$, been related to a dS space ($\lambda> 0$) or
AdS space ($\lambda< 0$) via Einstein's equation,
$R_{\mu\nu}-\frac{g_{\mu\nu}}2\,R -\lambda\,g_{\mu\nu}=0 $, where
Riemann and Ricci tensors are
\begin{eqnarray}
R_{\lambda \mu\nu \rho}=\frac R6 \,(g_{\lambda\nu}g_{\mu
\rho}-g_{\lambda\rho}g_{\mu\nu})
 \, \, ,\,\,\,\,\,R_{ \mu\nu }=\frac R3 \,g_{\mu\nu}\,\,, \label{eq30}
\end{eqnarray}
respectively, and
\begin{eqnarray}
R=-6 \lambda\,\,. \label{eq30b}
\end{eqnarray}
(\ref{eq22}) is
\begin{eqnarray}
\Omega^{\alpha \beta \sigma \lambda } =
 M^2\,(g^{\alpha \beta }g^{\sigma \lambda}
- g^{\sigma \beta }g^{\alpha \lambda } )
 \, \, , \label{eq31}
 \end{eqnarray}
where
\begin{eqnarray}
M^2 = m^2+\sigma R\, \, , \label{eq31b}
 \end{eqnarray}
with $\sigma
\equiv \frac{2}{3}\,a - b$. Using (\ref{eq31}), the action (\ref{eq13}) takes the form
\begin{equation}
S_{sd\lambda}=\int \frac{d^3 x}{2} \sqrt {-g}( m\,\varepsilon
^{\mu \nu \lambda }{h_\mu }^\alpha \nabla _\nu h_{\lambda \alpha
}-M^2\,(h_{\mu \nu }h^{\nu \mu }-h^2 ) ) \, \, , \label{eq32}
\end{equation}
and with the help of (\ref{eq24}) and  (\ref{eq31}), the object
$\mathcal{N}^{ij}$  becomes
\begin{equation}
\mathcal{N}^{ij } \equiv \bigg(\frac{6M^4-Rm^2}m \bigg)
\,\varepsilon ^{ij} \, \, . \label{eq33}
\end{equation}
The consistence relation (\ref{eq29}) is now
\begin{equation}
6M^4-Rm^2\neq 0\, \, . \label{eq34}
\end{equation}
Considering (\ref{eq31b}) we can think about this relation as a
restriction on $m^2$ in terms of scalar curvature and $\sigma$.
This means
\begin{equation}
m^2\neq {m_{\pm}}^2 \equiv \frac {R}{12}\,\big(1-12\sigma \pm
\sqrt{1-24\sigma }\,\big)
 \, \, , \label{eq35}
\end{equation}
showing the existence of some forbidden mass values in order to
have consistency, which represents a well known fact in context of
higher spin theories\cite{Hig}.

Now, lagrangian constraints are revisited, this time in  dS/AdS
space. The primary nine, (\ref{eq15}) are
\begin{equation}
\Phi ^{(1) \mu \alpha }\equiv m\,\varepsilon ^{\mu \nu \lambda }
\nabla _\nu {h_\lambda}^\alpha + M^2\,(g^{\mu \alpha }h -
h^{\alpha \mu }) \approx 0 \, \, . \label{eq36}
\end{equation}
Next, secondary constraints (\ref{eq16})
\begin{equation}
\Phi ^{(2) \alpha }\approx M^2\,(\nabla ^{\alpha}h
 -\nabla _{\mu}h^{\alpha \mu })+\frac{mR}{6}\, \varepsilon ^{\alpha
 \sigma
\lambda} h_{\sigma \lambda}
 \approx  0\, \, , \label{eq37}
\end{equation}
which are written as $\Phi ^{(2) \alpha }\approx \nabla _\mu \Phi
^{(1) \mu \alpha }- \frac{M^2}{m}\,\varepsilon ^{\alpha \sigma
\lambda }{\Phi ^{(1)}}_{\mu \alpha
}=\big(\frac{m^2R-6M^4}{6m}\big)\,\varepsilon ^{\alpha \sigma
\lambda} h_{\sigma \lambda}\approx 0$. So, the symmetry property
for the $h_{\mu \nu}$ field, in virtue of (\ref{eq34}), is gained.

Preservation of $\Phi ^{(2) \alpha }\approx 0$ provides three
more constraints
\begin{equation}
\Phi ^{(3) \alpha }\approx
\bigg(\frac{m^2R-6M^4}{6m}\bigg)\,\varepsilon ^{\alpha
 \sigma
\lambda}\nabla _0 h_{\sigma \lambda}\approx  0\, \, . \label{eq38}
\end{equation}
Its temporal component, $\Phi ^{(3) 0 }\approx 0$ is expressed as
\begin{equation}
\Phi ^{(3) 0 }\approx \nabla _\mu \Phi ^{(2) \mu
}+\bigg(\frac{6M^4-m^2R}{6m^2}\bigg)\, {\Phi ^{(1) \mu}}_\mu =
\frac{M^2}{3m^2}\,(6M^4-m^2R)h \approx 0\, \, . \label{eq39}
\end{equation}
It says that the spin 2 field is traceless  (obviously if $M^2
\neq 0$ ). The last constraint arise from the preservation of
$\Phi ^{(3) 0 }\approx 0$
\begin{equation}
\Phi ^{(4) }\equiv \nabla _0 \Phi ^{(3) 0 }\approx
\frac{M^2}{3m^2}\,(6M^4-m^2R)\nabla _0h \approx 0\, \, .
\label{eq40}
\end{equation}

We observe that traceless and transverse properties for a
consistent description of the selfdual field, demands the
additional condition
\begin{equation}
M^2\neq  0\, \, , \label{eq41}
\end{equation}
as a consequence of  equations (\ref{eq37}), (\ref{eq39}) and
(\ref{eq40}). If one relax this restriction on $M^2$ (i.e., $M^2=
0$), the lagrangian system will not furnish the expected number of
degrees of freedom.

Imposing (\ref{eq41}) we can construct a quadratical
Klein-Gordon-like field equation  for ${h^{(s)Tt}}_{\mu \nu}$, as
follows
\begin{equation}
\big( \Box{} -\frac{M^4}{m^2}+\frac{R}{2} \big){h^{(s)Tt}}_{ \mu
\nu}=0 \, \, . \label{eq42}
\end{equation}
with $\Box{}\equiv \nabla _\alpha \nabla ^\alpha$. This equation
is clearly hyperbolic and causal, because we can rewrite it in the
form
\begin{equation}
({\mathcal{M}^{\beta\sigma}}_{\rho\alpha})^{\mu\nu}\nabla _\mu
\nabla _\nu{h^{(s)Tt}}_{\beta \sigma}+...=0\, \, . \label{eq43}
\end{equation}
 where
$({\mathcal{M}^{\beta\sigma}}_{\rho\alpha})^{\mu\nu}=g^{\mu\nu}{\delta^{\beta\sigma}}_
{\rho\alpha}$ and ${\delta ^{\beta\sigma}}_{\rho \alpha}\equiv
\frac{1}{2}({\delta^{\beta}}_{\rho}{\delta^{\sigma}}_{\alpha}+
{\delta^{\sigma}}_{\rho}{\delta^{\beta}}_{\alpha})$. Then, with
the help of the three-vectors  $n_\mu$, we define the
characteristic matrix
\begin{eqnarray}
{\mathcal{M}^{\beta\sigma}}_{\rho\alpha}(n)={\delta
^{\beta\sigma}}_{\rho \alpha}\,n^2\, \, , \label{eq44}
\end{eqnarray}
whose characteristic equation is
\begin{eqnarray}
det(\mathcal{M})=(n^2)^6=0 \, \, , \label{eq45}
\end{eqnarray}
which has a null vector solution.

As dS/AdS space are conformally flat, its light cones are
equivalent to those of Minkowski (i.e., they are related through a
Weyl map), and we can write $n^2=0$ in a locally Weyl-flat frame
(using the fact that the conformal transformation for metric is
$g_{\mu\nu}=\Omega ^2 \eta_{\mu\nu}$), as follows
\begin{eqnarray}
-(n_0)^2+n_in_i=0 \, \, , \label{eq46}
\end{eqnarray}
which describes an hyperbolic ($n_0$ is real) and causal ($n^2=0$
implies there is not time-like three-vectors) propagation.

On the other hand, the selfdual massive theory studied holds
forbidden mass values in dS/AdS spaces because of (\ref{eq41})
(i.e., $m^2 \neq -\sigma R$). These values can be resumed as
follows

\begin{center}
\begin{tabular}{|c|c|c|}
 \hline
\textbf { $R$} & \textbf{ $\sigma$} & \textbf{ $forbidden\,\,\,m$} \\
\hline \hline
$>0 \,(AdS)$& $0 \leq \sigma \leq \frac{1}{24}$ & $m_{\pm}$  \\
\hline
$>0 \,(AdS)$& $<0$ & $\sqrt{-\sigma R}$  \\
\hline
$<0 \,(dS)$ & $>0$ & $\sqrt{-\sigma R}$  \\
\hline

\end{tabular}
\vspace{0.5cm}

% table 1

\end{center}

We note that, in the study of the selfdual massive spin 2 field
theory coupled with gravity a well known fact is verified: inside
a possible set of solutions, those of constant curvature spaces
respect the number of degrees of freedom and causality. However,
in contrast to other type of spin 2 theories\cite{BGP}, the
selfdual massive one does not have massless limit, and $M^2\neq 0$
is demanded in order to guarantee equivalence between constraint
system and symmetry, traceless and transverse properties of
selfdual massive field with an hyperbolic and causal equation
provided.

There are other issues related with the condition $M^2\neq 0$. On
one hand, this condition, in a dS/AdS background mantains the
selfdual lagrangian, (\ref{eq32}), conformally variant due to the
non null trace of the energy-momentun associated with the selfdual
field, ${T^\mu}_\mu=-\frac{M^2}2\,{h^{(s)Tt}}_{\mu \nu}h^{(s)Tt\mu
\nu}$.

Moreover, the critical value $M^2=0$ reveals the existence of an
expected discontinuity in the degrees of freedom´s count, because
it gives rise to a non consistent set of lagrangian constraints,
which is associated with the nature of the quadratical terms in
the action, and not to the features of the gravitational
interaction. This kind of discontinuity can be illustrated from in
the flat space model, as follows. Let us consider the two
parameter action
\begin{eqnarray}
{S_{m_1,\,m_2}}= \int d^3 x (\frac{m_1}2\,\epsilon ^{\mu \nu
\lambda }{h_\mu }^\alpha
\partial _\nu
h_{\lambda \alpha }-\frac{{m_2}^2}2(h_{\mu \nu }h^{\nu \mu }-h^2
)) \, \, ,\label{eq47}
\end{eqnarray}
which reproduces the selfdual massive model when we choose
$m_1=m_2=m$. Particularly, when $m_2=0$, the action (\ref{eq47})
describes a model with no degrees of freedom (in fact, the reduced
action becomes null identically). However, if $m_2\neq 0$ is
considered during the procedure that lead us to the reduced
action, one can arrive to the expected relation:
${S_{m_1,\,m_2}}^{*}=\int d^3 x \{P\dot{Q}-\frac12 P^2 + \frac12
\,Q(\Delta -\mathbb{M}^2)Q\}$, whith $\mathbb{M}\equiv
\frac{{m_2}^2}{m_1}$\,,\,\, $P\equiv \sqrt 2 \,m_2 \Delta
{h^{(s)TL}}$ and $Q\equiv \sqrt 2 \,\frac{m_1}{m_2} \Delta
{h^{(s)TT}}$ , which is a singular function at $m_2=0$, saying
that the model (\ref{eq47}) does not have a well defined limit at
$m_2=0$.

In an analogous way, a discontinuity does appear in  the selfdual
massive model when we consider a dS/AdS background, (\ref{eq32})
at the critical value $M^2 = 0$. In fact, this behavior is
manifest if we observe that equation (\ref{eq36}) is now gauge
invariant under
\begin{equation}
\delta {h^{(s)Tt}}_{\mu\nu }=(\nabla_\mu \nabla_\nu
-\frac{R}6\,g_{\mu \nu })\omega (x) \, \, , \label{eq48}
\end{equation}
which says that the only degree of freedom due to
${h^{(s)Tt}}_{\mu\nu }$, can be gauged away and the theory with
$M^2 = 0$ does not propagating degrees of freedom as in the flat
case.

%============================================================================================

If in the context of a curved space-time, we want to realize a
procedure in order to obtain a reduced action for selfdual massive
spin 2 theory, and then description of the only propagated degree
of freedom through a field like ${{h^{(s)Tt}}_{\mu \nu }}^\pm$. In
the flat case it can be seen that the symbol ``$\pm$'' is
associated with a propagation of spin $\pm 2$\cite{Arias}. In a
curved space-time this ''flat'' procedure will find serious
obstacles. Essentially, this is related with the problem of the
Fourier transform in curved spaces\cite{r57} and the definition of
arbitrary powers of D'Alembertian, and as a consecuence of the
obscure business to obtain projectors.

However, we can say something following a covariant procedure in
order to obtain the reduced action, starting with the a
symmetric-antisymmetric decomposition
\begin{equation}
h_{\mu \nu }\equiv { h^{(s)}}_{\mu \nu }+\varepsilon _{\mu \nu
\lambda }V^{\lambda}\, \, . \label{r42}
\end{equation}
Using this in (\ref{eq32}), conduce us to
\begin{eqnarray}
S_{sd\lambda}= \int d^3 x \,\sqrt{-g}\big(\,\frac m2\,\varepsilon
^{\mu \nu \sigma }g^{\beta\alpha}{h^{(s)}}_{\mu \beta} \nabla _\nu
{h^{(s)}}_{\sigma \alpha }-\frac {M^2}{2}\,({h^{(s)}}_{\mu \nu
}h^{(s){\mu \nu }}-{h^{(s)}}^2 )+\nonumber
\\+\, mV^\mu (\nabla _\mu {h^{(s)}} -\nabla
_\nu{{h^{(s)}}_\mu}^\nu)-\frac m2\,\varepsilon ^{\mu \nu \sigma
}V_\mu \nabla _\nu V_\sigma -M^2V_\mu V^\mu \big) \, \, ,\nonumber
\\\label{r43}
\end{eqnarray}
and the field equations
\begin{eqnarray}
m\varepsilon ^{\mu \nu \lambda } \nabla_\mu{{h^{(s)}}_\nu }^\alpha
+m\varepsilon ^{\mu \nu \alpha }\nabla_\mu{{h^{(s)}}_\nu }^\lambda
-2M^2 h^{(s)\lambda
\alpha }+  2M^2 g^{\lambda \alpha }h^{(s)}+\nonumber \\
-2mg^{\lambda \alpha }\nabla_\mu V^\mu + m(\nabla^\lambda
V^\alpha+ \nabla^\alpha V^\lambda)=0
 \, \, , \label{rb1}
\end{eqnarray}
\begin{equation}
m\varepsilon ^{\mu \nu \lambda }\nabla_\mu V_\nu +2M^2 V^\lambda
-m\nabla^\lambda h^{(s)}+ m\nabla_\mu h^{(s) \mu\lambda }=0
 \, \, . \label{rb2}
\end{equation}
The trace and divergence of (\ref{rb1}) give
\begin{eqnarray}
 M^2h^{(s)}-m\nabla_\mu V^\mu =0
 \, \, , \label{rb3}
\end{eqnarray}
\begin{eqnarray}
m\varepsilon ^{\mu \nu \lambda } \nabla_\nu
{\mathcal{H}}_\lambda -2M^2{\mathcal{H}}\,^\mu+
 2M^2 \nabla^\mu h^{(s)} +m\Box{}V^\mu +\nonumber \\-\,
m\nabla^\mu \nabla_\alpha V^\alpha-\frac{mR}{3}\,V^\mu=0
 \, \, , \label{rb4}
\end{eqnarray}
with the notation ${\mathcal{H}}_\lambda \equiv
\nabla_\alpha{{h^{(s)}}_\lambda }^\alpha$. Using (\ref{rb4}) in
(\ref{rb2}), we get
\begin{equation}
(Rm^2-6M^4)V_\sigma =0.
 \, \, , \label{rb5}
\end{equation}
Taking into account the restriction (\ref{eq34}), we get $V_\sigma
=0$. This last relation with (\ref{rb3}) gives the suplementary
$h^{(s)}=0$, and equation (\ref{rb2}) assure
${\mathcal{H}}_\lambda \equiv\nabla_\alpha{{h^{(s)}}_\lambda
}^\alpha =0$. Then, it is confirmed that in constant curvature
spaces the selfdual massive spin 2 theory is described by a
symmetric-transverse-traceless field, ${h^{(s)Tt}}_{\mu \nu }$,
and the reduced action will take the form
\begin{eqnarray}
{S_{sd\lambda}}^{(2)*}= \int d^3 x\,\sqrt{-g}\, \big(\,\frac
{m}2\,\varepsilon ^{\mu \nu \sigma }{{h^{(s)Tt}}_\mu }^\alpha
\nabla _\nu {h^{(s)Tt}}_{\sigma \alpha }-\frac {M^2}2\,
{h^{(s)Tt}}_{\mu \nu }h^{(s)Tt\mu \nu } \big) \, \, , \label{r65}
\end{eqnarray}
which the equations of motion
\begin{eqnarray}
m\,\varepsilon ^{\sigma\mu \nu } \nabla _\mu{{h^{(s)Tt}}_\nu
}^\beta -M^2 h^{(s)Tt\sigma \beta }=0
 \, \, . \label{r66}
\end{eqnarray}
From this, the causal propagation (\ref{eq42}) is obtained.

At a point $p$ of the manifold $\mathcal{M}$, it can be attached a
tangent space, $T_p(\mathcal{M})$ with locally coordinates $\xi^a$
provided. So, in this reference the hyperbolic-causal equation is
\begin{equation}
\big( \Box{}_{(\xi)} -\frac{M^4}{m^2}+\frac{R}{2}
\big){h^{(s)Tt}}_{ ab}(\xi)=0 \, \, , \label{r67}
\end{equation}
with $\Box{}_{(\xi)}\equiv \partial^a \partial_a$. Next, we define
the locally ''+'' and ''$-$'' parts of ${h^{(s)Tt}}_{ab}(\xi)$ in
the way
\begin{equation}
{{{h^{(s)Tt}}}_{ab}}^\pm \equiv \frac{1}{2}\bigg(\frac{{\delta
^d}_a{\delta ^c}_b}{q}  \pm {\delta ^d }_a {\epsilon _b}^{rc
}\frac{\partial _r
}{{\Box{}_{(\xi)}}^{\frac{1}{2}}}\bigg){{h^{(s)Tt}}}_{dc}
 \, \, , \label{r68}
\end{equation}
where the parameter $q \equiv \sqrt{ 1- \frac{Rm^2}{2M^4}}$. Then,
with the local on-shell relation (\ref{r67}),  it can be obtained
that
\begin{equation}
\big( \Box{}_{(\xi)} -\frac{M^4}{m^2}+\frac{R}{2}
\big){{{h^{(s)Tt}}}_{ab}}^\pm(\xi)=0 \, \, , \label{r69}
\end{equation}
\begin{equation}
{{{h^{(s)Tt}}}_{ab}}^\mp(\xi)=0 \, \, , \label{r70}
\end{equation}
saying that the only degree of freedom locally propagated is
described through ${{{h^{(s)Tt}}}_{ab}}^+$
(${{{h^{(s)Tt}}}_{ab}}^-$ ), if the spin is $+2$($-2$). It can be
observed that expression (\ref{r68}) can be rewritten as
${{{h^{(s)Tt}}}_{ab}}^\pm \equiv {P^{\pm dc
}}_{ab}{{h^{(s)Tt}}}_{dc}$, where
\begin{equation}
{P^{\pm dc }}_{ab}\equiv \frac{1}{4}\bigg(\frac{1}{q}\big({\delta
^d}_a{\delta ^c}_b+{\delta ^c}_a{\delta ^d}_b\big) \pm
\big({\delta ^d }_a {\epsilon _b}^{rc }+{\delta ^c }_a {\epsilon
_b}^{rd }\big)\,\frac{\partial _r
}{{\Box{}_{(\xi)}}^{\frac{1}{2}}}\bigg)
 \, \, , \label{r71}
\end{equation}
is not a projector (i.e., ${P^{\pm dc }}_{ab}{P^{\pm ab
}}_{ef}\neq {P^{\pm dc }}_{ef}$), because $q\neq 1$.

%============================================================================================

Finally, we examine the one-particle exchange amplitude which
describes the interaction between sources. This starts with the
selfdual massive spin 2 action in the form (\ref{r43}), but now
minimally coupled with an external, symmetric, conserved source
${T^{(s)}}_{\mu\nu}(x)$ (i.e., $\nabla^\mu
{T^{(s)}}_{\mu\nu}(x)=0$) as follows
\begin{eqnarray}
S_{sd\lambda s}= \int d^3 x \,\sqrt{-g}\big(\,\frac
m2\,\varepsilon ^{\mu \nu \sigma }g^{\beta\alpha}{h^{(s)}}_{\mu
\beta} \nabla _\nu {h^{(s)}}_{\sigma \alpha }-\frac
{M^2}{2}\,({h^{(s)}}_{\mu \nu }h^{(s){\mu \nu }}-{h^{(s)}}^2
)+\nonumber
\\+\, mV^\mu (\nabla _\mu {h^{(s)}} -\nabla
_\nu{{h^{(s)}}_\mu}^\nu)-\frac m2\,\varepsilon ^{\mu \nu \sigma
}V_\mu \nabla _\nu V_\sigma -M^2V_\mu V^\mu +\kappa {h^{(s)}}_{\mu
\nu}T^{(s)\mu\nu}\big) \, \, ,\nonumber
\\\label{ea1}
\end{eqnarray}
where $\kappa$ is a coupling parameter.

The field equations are emerging from (\ref{ea1}) are
\begin{eqnarray}
m\varepsilon ^{\mu \nu \lambda } \nabla_\mu{{h^{(s)}}_\nu }^\alpha
+m\varepsilon ^{\mu \nu \alpha }\nabla_\mu{{h^{(s)}}_\nu }^\lambda
-2M^2 h^{(s)\lambda
\alpha }+  2M^2 g^{\lambda \alpha }h^{(s)}+\nonumber \\
-2mg^{\lambda \alpha }\nabla_\mu V^\mu + m(\nabla^\lambda
V^\alpha+ \nabla^\alpha V^\lambda)=-2\kappa T^{(s)\lambda\alpha}
 \, \, , \label{ea2}
\end{eqnarray}
\begin{eqnarray}
m\varepsilon ^{\mu \nu \lambda }\nabla_\mu V_\nu +2M^2 V^\lambda
-m\nabla^\lambda h^{(s)}+ m\nabla_\mu h^{(s) \mu\lambda }=0
 \, \, . \label{ea3}
\end{eqnarray}
Divergence and trace of (\ref{ea2}) give
\begin{eqnarray}
\varepsilon ^{\mu \nu \lambda }
\nabla_\nu\nabla_\alpha{{h^{(s)}}_\lambda }^\alpha -\frac{2M^2}{m}
\nabla_\alpha h^{(s)\mu \alpha }+  \frac{2M^2}{m} \nabla^\mu
h^{(s)} -2m\nabla^\mu\nabla_\lambda V^\lambda +\nonumber \\+
\nabla_\lambda \nabla^\mu V^\lambda+\Box{}V^\mu=0
 \, \, , \label{e2}
\end{eqnarray}
\begin{equation}
2M^2h^{(s)} -2m\nabla_\mu V^\mu =-\kappa T^{(s)}
 \, \, . \label{e3}
\end{equation}
Curl of (\ref{ea3}) is
\begin{equation}
-{\varepsilon_\sigma }^{\rho \mu }
\nabla_\rho\nabla_\nu{{h^{(s)}}_\mu }^\nu
-\Box{}V_\sigma+\nabla_\lambda \nabla_\sigma V^\lambda
-\frac{2M^2}{m} \varepsilon_{\sigma \rho \mu }\nabla^\rho V^\mu=0
 \, \, , \label{e4}
\end{equation}
and with the help of (\ref{e2}), conduce us to
\begin{equation}
(m^2R-6M^4)V_\sigma=0
 \, \, , \label{e5}
\end{equation}
which again says that $V_\sigma =0$. Then we can rewrite
(\ref{ea2}), (\ref{ea3}) and (\ref{e3}) as follows
\begin{equation}
m\varepsilon ^{\mu \nu \lambda } \nabla_\mu{{h^{(s)}}_\nu }^\alpha
+m\varepsilon ^{\mu \nu \alpha }\nabla_\mu{{h^{(s)}}_\nu }^\lambda
-2M^2 h^{(s)\lambda \alpha }+  2M^2 g^{\lambda \alpha }h^{(s)}
=-2\kappa T^{(s)\lambda\alpha}
 \, \, , \label{e6}
\end{equation}
\begin{equation}
\nabla^\lambda h^{(s)}-\nabla_\mu h^{(s) \mu\lambda }=0
 \, \, , \label{e7}
\end{equation}
\begin{equation}
2M^2h^{(s)}=-\kappa T^{(s)}
 \, \, . \label{e8}
\end{equation}

For the computation of the exchange amplitude we need the
decomposition
\begin{equation}
{h^{(s)}}_{\mu\nu}={h^{(s)Tt}}_{\mu\nu}+\nabla_\mu
{a^T}_\nu+\nabla_\nu {a^T}_\mu+\nabla_\mu \nabla_\nu \phi +
g_{\mu\nu}\psi
 \, \, , \label{e9}
\end{equation}
where $\nabla^\mu {a^T}_\mu =0$. The following relations arise
from (\ref{e9})
\begin{equation}
h^{(s)}=\Box{} \phi + 3\psi
 \, \, , \label{e10}
\end{equation}
\begin{equation}
(-\Box{}+\frac{R}{3}) {a^T}_\mu+\frac{R}{3}\nabla_\mu \phi +
2\nabla_\mu \psi=0
 \, \, , \label{e11}
\end{equation}
where the last one is obtained with the help of (\ref{e7}).
Divergence of (\ref{e11}) provides $R\Box{}\phi+6\Box{}\psi=0$,
and using this in (\ref{e10}) with (\ref{e8}), we get
\begin{equation}
(\Box{}-\frac{R}{2})\psi=\frac{R\kappa}{12M^2} T^{(s)}
 \, \, . \label{e12}
\end{equation}

Now we need to write down ${h^{(s)Tt}}_{\mu\nu}$ in terms of the
source. The Tt part of (\ref{e6}) is
\begin{equation}
m\varepsilon ^{\mu \nu \lambda } \nabla_\nu{{h^{(s)Tt}}_\lambda
}^\alpha -M^2 h^{(s)Tt\mu \alpha }=-\kappa T^{(s)Tt\mu\alpha}
 \, \, , \label{e13}
\end{equation}
from which we obtain the hyperbolic-causal equation for
${h^{(s)Tt}}_{\mu\nu}$
\begin{equation}
\big( \Delta^{(2)} +\frac{M^4}{m^2}+\frac{R}{2}
\big)h^{(s)Tt\mu\alpha}=\frac{\kappa M^2}{m^2}T^{(s)Tt\mu\alpha}
+\frac{\kappa}{m}\,\varepsilon ^{\mu \rho \sigma }\nabla_\rho
{{T^{(s)Tt}}_\sigma}^\alpha\, \, , \label{e14}
\end{equation}
where $\Delta^{(2)}$ is the Lichnerowicz operator which obeys the
following properties\cite{Lich}
\begin{equation}
\Delta^{(0)}\phi=-\Box{}\phi\, \, , \label{e15a}
\end{equation}
\begin{equation}
\nabla^\mu\Delta^{(1)}V_\mu=\Delta^{(0)}\nabla^\mu V_\mu\, \, ,
\label{e15b}
\end{equation}
\begin{equation}
\Delta^{(2)}\nabla_{(\mu}V_{\nu )}=\nabla_{(\mu}\Delta^{(1)}V_{\nu
)}\, \, , \label{e15c}
\end{equation}
\begin{equation}
\nabla^\mu\Delta^{(2)}{h^{(s)}}_{\mu\nu}=\Delta^{(1)}\nabla^\mu
{h^{(s)}}_{\mu\nu}\, \, , \label{e15d}
\end{equation}
\begin{equation}
\Delta^{(2)}(g_{\mu\nu}\phi)=g_{\mu\nu}\Delta^{(0)}\phi\, \, ,
\label{e15e}
\end{equation}
and ${T^{(s)Tt}}_{\mu\nu}$ is given by
\begin{equation}
{T^{(s)Tt}}_{\mu\nu}={T^{(s)}}_{\mu\nu}-\frac{g_{\mu\nu}}{2}T^{(s)}+\frac{1}{2}
(\nabla_\mu \nabla_\nu-
\frac{R}{6}g_{\mu\nu})(\Box{}-\frac{R}{2})^{-1}T^{(s)}\, \, .
\label{e16}
\end{equation}

The exchange amplitude between two covariant conserved sources is
$A=\int d^3x\sqrt{-g} \mathcal{A}$, where $\mathcal{A}\equiv
{T'^{(s)}}_{\mu\nu}h^{(s)\mu\nu}$. Up to boundary terms we can
write $A$ using
\begin{equation}
\mathcal{A}= {T'^{(s)}}_{\mu\nu}h^{(s)Tt\mu\nu}+T'^{(s)}\psi \, \,
.\label{e17}
\end{equation}
Considering (\ref{e12}), (\ref{e14}) and (\ref{e16}), we obtain
\begin{eqnarray}
\frac{\mathcal{A}}{\kappa}=
\frac{M^2}{m^2}\,\,{T'^{(s)}}_{\alpha\beta}(\Delta^{(2)}+\mu^2)^{-1}T^{(s)\alpha\beta}
-\frac{M^2}{2m^2}\,\,T'^{(s)}(-\Box{}+\mu^2)^{-1}T^{(s)}+\nonumber\\
+\frac{2}{m}\,\,{T'^{(s)}}_{\alpha\beta}(\Delta^{(2)}+\mu^2)^{-1}\varepsilon
^{(\alpha \rho \sigma }\nabla_\rho {T^{(s)Tt\beta )}}_\sigma
-\frac{R}{12m^2}\,\,T'^{(s)}(-\Box{}+\frac{R}{2})^{-1}T^{(s)}+\nonumber\\
+\frac{M^2R}{12m^2}\,\,T'^{(s)}(-\Box{}+\mu^2)^{-1}(-\Box{}+\frac{R}{2})^{-1}T^{(s)}
\, \, ,\label{e18}
\end{eqnarray}
where $\mu^2\equiv \frac{M^4}{m^2}+\frac{R}{2}$.

In the flat limit, looking at the first three terms in
(\ref{e18}), two of them will be proportional to $\frac{M^2}{m^2}$
and the other to $\frac{2}{m}$. They correspond to the amplitude
of a massive selfdual massive spin 2 in 2+1 dimensions. For the
remaining terms they give a cosmological contribution which
disappears in the flat limit. In the curved case it can be
observed that these last terms have an unphysical pole at
$\Box{}=\frac{R}{2}$ which do not propagate (i.e., the residue in
the amplitude is
$\frac{M^2R}{12m^2}(-\frac{R}{2}+\mu^2)^{-1}-\frac{R}{12M^2}=0$,
whatever the sign of the cosmological constant). On the other
hand, the physical pole $\Box{}=\mu^2$ has the residue
\begin{equation}
\mathcal{R}_{(\Box{}=\mu^2)}=-\kappa\frac{M^2}{2m^2}\big(
1-\frac{\lambda m^2}{M^4}\big) \, \, ,\label{e19}
\end{equation}
which is clearly non null in an AdS space-time.

As a concluding remark, the selfdual massive spin 2 model in
dS/AdS background exhibits frobidden mass values in order to
guarantee consistency, and they are given by (\ref{eq34}) and
(\ref{eq41})
\begin{equation}
6M^4-Rm^2\neq 0\, \, , \label{eq49a}
\end{equation}
\begin{equation}
M^2\neq  0\, \, , \label{eq49b}
\end{equation}
with $M^2=m^2+\sigma R$. Both contain information about the
background and they match in the flat space-time limit with the
consistence condition of the selfdual massive model: $m\neq 0$.
Moreover, $M^2$ does appear as a quadratical power of a ''mass''
in action (\ref{eq32}), this can be thought as the curvilinear
version of the two parameters flat action, (\ref{eq47})(which
contains the flat selfdual massive spin 2 model). So, the presence
of $M^2\neq 0$ in the action (\ref{eq32}), guarantees a
conformally variant selfdual massive model in dS/AdS, matching
with the same situation in flat theory. However, one can
distinguish between dS or AdS because the sign of the residue,
$\mathcal{R}_{(\Box{}=\mu^2)}$ is sensitive when $\lambda
> 0$ (dS).

\vspace{5pt}

\noindent{\bf Aknowlegments}

This work is partially supported by  projects: G-2001000712 of
FONACIT, PI-03-00-5753-04 of CDCH-UCV and 1102-06 of CDCH-UC.


\begin{thebibliography}{6}
\bibitem{ArDe}   C. Aragone, S. Deser, {\it Il Nuovo Cim. }{\bf 3A}, 709 (1971);
{\it Il Nuovo Cim. }{\bf 57B}, 33 (1980).
\bibitem{Hig} A. Higuchi, {\it Nuc. Phys. }{\bf B282}, 397 (1987);
{\it Nuc. Phys. }{\bf B325}, 745 (1989); {\it Class. Q. Grav.
}{\bf 6}, 397 (1989).
\bibitem{DPS} M. J. Duff, C. N. Pope, K. S. Stelle, {\it Phys. Lett. }{\bf B223}, 386 (1989).
\bibitem{Hohm:2005sc} O.~Hohm, Phys.\ Rev.\  D {\bf 73}, 044003 (2006) [arXiv:hep-th/0511165];
O.~Hohm, Class.\ Quant.\ Grav.\  {\bf 24}, 2825 (2007)
[arXiv:hep-th/0611347]..
\bibitem{Ben} I. Bengtsson, {\it J. Math. Phys. }{\bf 36}, 5805 (1995)[arXiv:gr-qc/9411057].
\bibitem{HOD} A. Hindawi, B. A. Ovrut, D. Waldram, {\it Phys. Rev. }{\bf D53}, 5583 (1996)[arXiv:hep-th/9509147].
\bibitem{Kli} S. M. Klishevich, {\it Class. Q. Grav. }{\bf 16}, 2915 (1999)[arXiv:hep-th/9812005].
\bibitem{FrV} E. S. Fradkin, M. A. Vasiliev, {\it Nuc. Phys. }{\bf B291}, 141 (1987);
  M.A. Vasilev, {\it Phys. Lett. }{\bf B243}, 378 (1990).
\bibitem{Vas} M. A. Vasiliev, {\it Int. J. Mod. Phys. }{\bf D5}, 763 (1996)[arXiv:hep-th/9611024].
\bibitem{CuW} C. Cutler, R. M. Wald, {\it Class. Q. Grav. }{\bf 4}, 1267 (1987);
 R.M. Wald, {\it Class. Q. Grav. }{\bf 4}, 1279 (1987).
\bibitem{BGP} I. L. Buchbinder, D. M. Gitman, V. D. Pershin, {\it Phys. Lett. }{\bf
B492}, 161  (2000).
\bibitem{FiPa} M. Fierz, W. Pauli,  {\it Proc. Royal Soc. }{\bf A173}, 211 (1939).
\bibitem{CSH} S. J. Chang, {\it Phys. Rev. }{\bf 161}, 1308 (1967);
L.P.S. Singh, C. R. Hagen, {\it Phys. Rev. }{\bf D9}, 898 (1974).
\bibitem{VeZ} G. Velo, D. Zwanziger, {\it Phys. Rev. }{\bf 188}, 2218 (1969);
 G. Velo, {\it Nuc. Phys. }{\bf B43}, 389 (1972).
\bibitem{Zwa} D. Zwanziger, {\it Lecture notes in physics  }{\bf 73}, 143 (1978).
\bibitem{GTy} D. M. Gitman, I. V. Tyutin, {\it Quantization of Fields with
Constraints} (Springer-Verlag, 1990).
\bibitem{AD}   C. Aragone, A. Khoudeir, {\it Phys. Lett. }{\bf B173}, 141 (1986).
\bibitem{DeserAD} S.Deser and J.G.McCarthy,{\it Phys. Lett. }{\bf B246}, 441 (1990)
  [Addendum-ibid. {\bf B248}, 473 (1990)]
\bibitem{G} R. Gaitan, {\it Sobre el problema del acoplamiento
de campos de espines altos en dimensi\'on $2+1$}, Doctoral Thesis,
Uiversidad Central de Venezuela, (2005).
\bibitem{PJA} P.J.Arias, {\it Spin 2 in (2+1)-dimensions},
Doctoral Thesis, Universidad Sim\' on Bol\'{\i}var, (In
spanish)(1994) [arXiv:gr-qc/9803083].
\bibitem{Arias}P.J.Arias and R.~Gaitan D., Rev.\ Mex.\ Fis.\  {\bf 52S3},
140 (2006) [arXiv:hep-th/0401107].
\bibitem{r57} S. Deser, {\it Ann. Inst. Henri Poincar\'e} {\bf VII}, 149 (1967).
\bibitem{Lich} A. Lichnerowicz, ''Propagateurs et commutateurs en
relativit\'e g\'enerale'', Institute de Hautes Etudes
Scientifiques, Publications Math\'ematiques, 10 (1961).



\end{thebibliography}
\end{document}